%% file: dpf2013-abuzayyad-paper.tex
\documentclass[12pt]{article}
\usepackage{graphicx}

\newcommand\pubnumber{DPF2013-205}
\newcommand\pubdate{September 30, 2013}

\def\utah{Department of Physics and Astronomy\\
University of Utah, Salt Lake City, UT 84112, USA}
\def\support{\footnote{Work supported by the National Science Foundation, Grant Number PHY-1069286}}

\def\Title#1{\begin{center} {\Large #1 } \end{center}}
\def\Author#1{\begin{center}{ \sc #1} \end{center}}
\def\Address#1{\begin{center}{ \it #1} \end{center}}

\newcommand\pubblock{\rightline{\begin{tabular}{l} \pubnumber\\
         \pubdate  \end{tabular}}}
\newenvironment{Abstract}{\begin{quotation}  }{\end{quotation}}
\newenvironment{Presented}{\begin{quotation} \begin{center} 
             PRESENTED AT\end{center}\bigskip 
      \begin{center}\begin{large}}{\end{large}\end{center} \end{quotation}}

\input econfmacros.tex

\begin{document}
\begin{titlepage}
\pubblock

\vfill
\Title{Cerenkov Events Seen by The TALE Air Fluorescence Detector}
\vfill
\Author{ Tareq Abu-Zayyad, Telescope-Array Collaboration\support}
\Address{\utah}
\vfill
\begin{Abstract}
The Telescope Array Low-Energy Extension (TALE) is a hybrid, Air Fluorescence Detector (FD) / Scintillator Array, designed to study cosmic ray initiated showers at energies above $\sim3\times10^{16}$ eV. Located in the western Utah desert, the TALE FD is comprised of 10 telescopes which cover the elevation range 31-58$^{\circ}$ in addition to 14 telescopes with elevation coverage of 3-31$^{\circ}$.
  As with all other FD's, a subset of the shower events recorded by TALE are ones for which the Cerenkov light produced by the shower particles dominates the total observed light signal.  In fact, for the telescopes with higher elevation coverage, low energy Cerenkov events form the vast majority of triggered cosmic ray events.  In the typical FD data analysis procedure, this subset of events is discarded and only events for which the majority of signal photons come from air fluorescence are kept.
  In this talk, I will report on a study to reconstruct the ``Cerenkov Events'' seen by the high elevation viewing telescopes of TALE.  Monte Carlo studies and a first look at real events observed by TALE look very promising.  Even as a monocular detector, the geometrical reconstruction method employed in this analysis allows for a pointing accuracy on the order of a degree.  Preliminary Monte Carlo studies indicate that, the expected energy resolution is better than 25$\%$.  It may be possible to extend the low energy reach of TALE to below $10^{16}$ eV.  This would be the first time a detector designed specifically as an air fluorescence detector is used as an imaging Cerenkov detector.
\end{Abstract}
\vfill
\begin{Presented}
DPF 2013\\
The Meeting of the American Physical Society\\
Division of Particles and Fields\\
Santa Cruz, California, August 13--17, 2013\\
\end{Presented}
\vfill
\end{titlepage}
\def\thefootnote{\fnsymbol{footnote}}
\setcounter{footnote}{0}

\section{Introduction}
\label{sec_intro}
The Telescope Array (TA) experiment was designed for the study of ultra high energy (above $\sim10^{18}$ eV) cosmic rays.  TA is the successor to the AGASA/HiRes experiments \cite{Teshima:1985vs,Sokolsky:2011zz} with the goal of improving on both.  TA is composed of three fluorescence detectors (FD's) \cite{AbuZayyad:2012qk,tafd-nim-a} and a large surface detector \cite{tasd-nim-a}.  TA is Located in Millard County, Utah, $\sim200$ km southwest of Salt Lake City.  The surface detector array is made up of 507 scintillation counters with 1.2 km spacing on a square grid.  The three fluorescence detectors have an elevation coverage of about $30^{\circ}$, and an azimuthal coverage of about $110^{\circ}$ overlooking the SD array.

  The TA Low Energy extension (TALE) detector \cite{tale_icrc2011} aims to lower the energy threshold of the experiment to well below $10^{17}$ eV.  This is mainly motivated by the interest in the galactic to extra-galactic transition in cosmic ray flux.

Located at the TA Middle Drum FD site, TALE adds an additional set of telescopes with high-elevation angle view to the site.  These complement the existing telescopes at Middle Drum.  In addition, an infill surface detector (SD) located closer to the FD site than the main TA array, and with closer spacing between the SD counters themselves, forms the second component of the ``hybrid detector''.  TALE operates as a hybrid detector (FD/SD) for best event quality in the intended range of operation, but can also operate as two separate detectors.  GPS timing allows for an observed cosmic ray shower (an event) observed separately by the FD and SD to be merged into a single event.  Events recorded by the FD which fail to trigger the SD, or if we choose to ignore the SD data, are referred to as monocular events.

As an air fluorescence detector, the TALE FD has an energy threshold of $\sim3\times10^{16}$ eV.  However, the detector is also sensitive to Cerenkov light produced by air showers, and if we think of the TALE FD as an Imaging Air Cerenkov Telescope (IACT) we find that we can extend the energy threshold of the detector down to $\sim3\times10^{15}$ eV, i.e. a full decade of energy lower than the original design goal of the experiment.

A study of the  $10^{16}$ eV decade is motivated by:
\begin{itemize}
\item{ the measurement of the spectrum using the same instrument used at higher energy.  This provides more overlap with km-scale detectors and better cross calibration of different detection techniques.}
\item{ measurement of the composition above the knee using the Xmax method.}
\item{ the measurement of the proton-air, nucleus-air cross section and comparison with accelerator measurements.}
\item{ possibly performing anisotropy and point source searches, as well as, gamma ray flux measurement.  We have {\em not} established whether or not this can be done, however, these topics are under investigation.}
\end{itemize}

In the following sections we first define the event data set we aim to study and contrast it with the event set that an FD analysis typically examines.  Next we give a brief description of the detector simulation procedure and explain the event reconstruction method used in this analysis.   Finally we discuss the performance of the method and provide an outlook for the future.

\section{Cerenkov Events Observed by TALE}
\label{sec:ckov_events}

\begin{figure}[htb]
\centering
\includegraphics[height=2.2in]{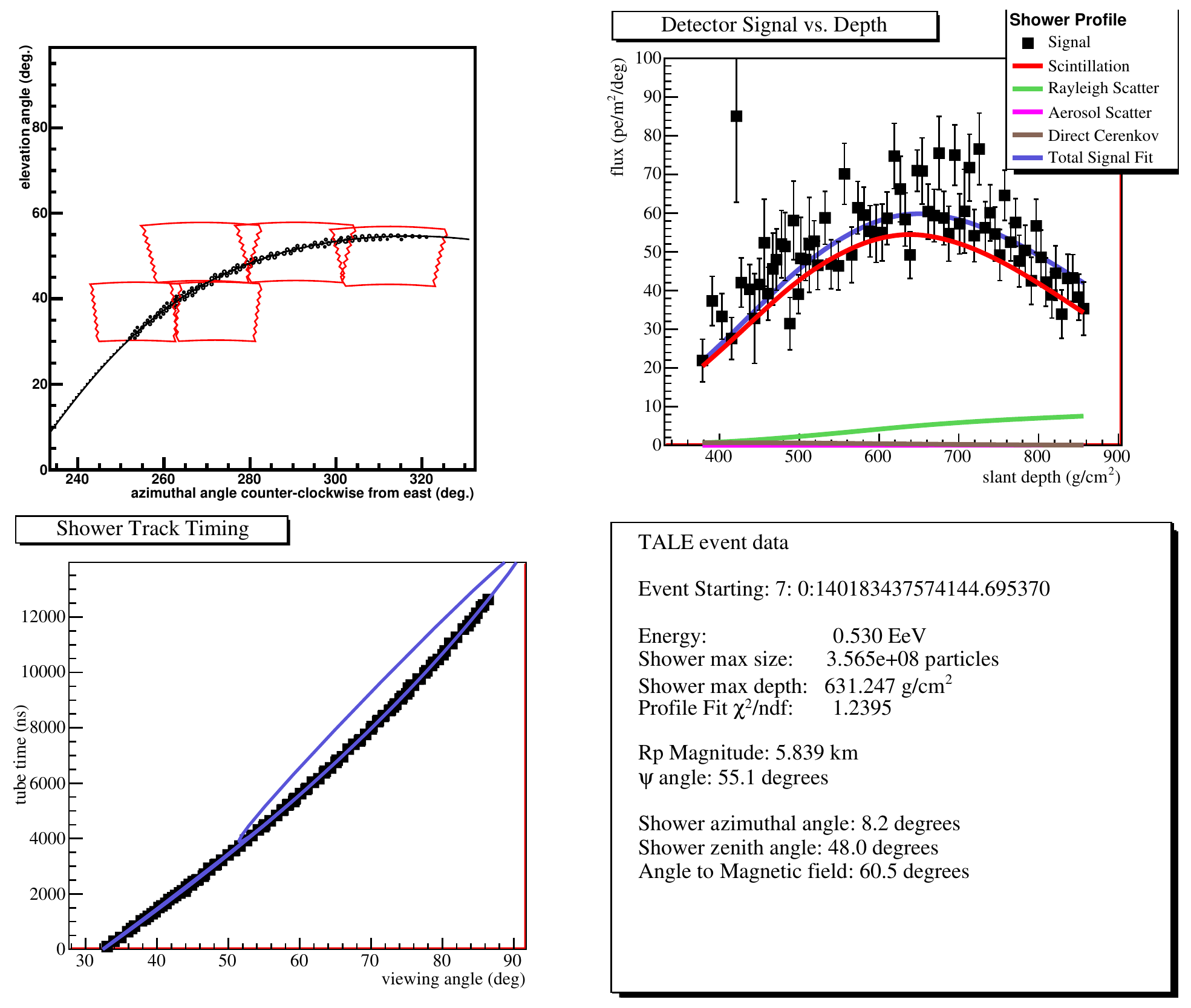}
\includegraphics[height=2.2in]{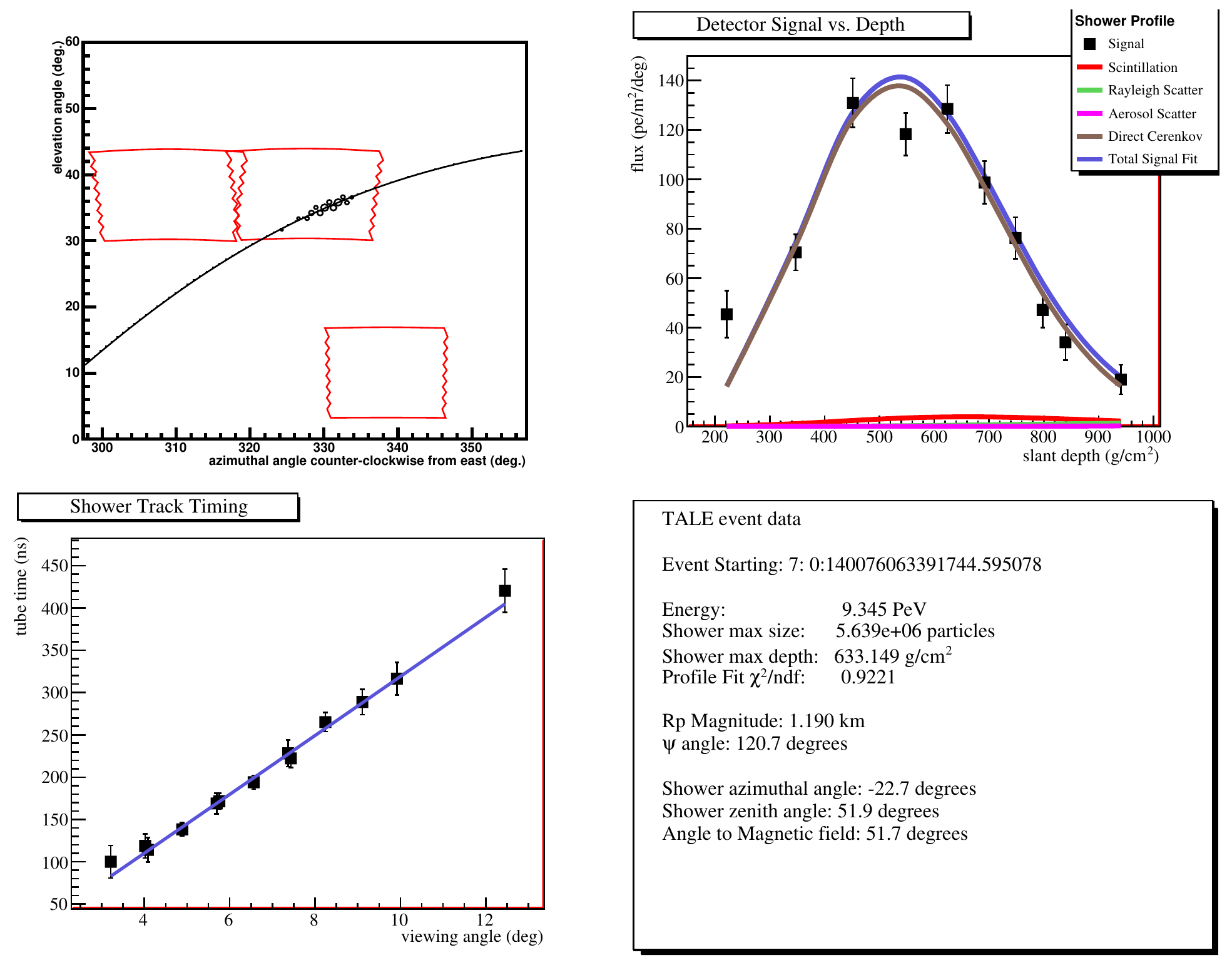}
\caption{Qualitative difference in observed Fluorescence events (left) and Cerenkov events (right).  Four panels per event show the event display (PMT trigger pattern), the time progression of triggerd PMT's, and reconstructed shower profile with relative contributions of FL/CL and scattered CL.} 
\label{fig:qualitative_events}
\end{figure}

The following is an overview of the qualitative differences between fluorescence and Cerenkov events, illustrated in Figure \ref{fig:qualitative_events}.  The light signal recorded by a fluorescence detector contains a contribution from Cerenkov light generated by the shower particles.  In event reconstruction, we distinguish among four contributes to the total observed light signal:
\begin{itemize}
  \item Direct Air Fluorescence light (FL).
  \item Direct Cerenkov light (CL).
  \item Rayleigh Scattered Cerenkov light (first order)
  \item Aerosols Scattered Cerenkov light (first order)
\end{itemize}
The relative contribution of the first two determines whether the event is counted as a Fluorescence or Cerenkov event.  The amount of scattered light is to be accounted for, however we seek events for which it is minimal.

Traditionally we require that FL constitute at least $\sim80\%$ of the received signal, although in some analyses this figure is relaxed to $60\%$.  In the new study described in this proceeding we require that the {\em Direct} CL contributes more than $80\%$ of the received signal.  This requirement insures good event reconstruction quality.  It also guarantees that we can divide the observed data into two distinct data sets, one Cerenkov and one fluorescence.  We can then perform a physics analysis, such as the energy spectrum measurement, using either set collected by a single detector and during the same time period, and compare the results in an overlap region as a systematic check on the results.

CL generated by a shower shares a trait with FL in that both are directly proportional to the number of shower particles for any given point in the shower development.  This property means that the observed CL signal can be used to infer the shower properties (energy and $x_{max}$) in a similar way to how the FL is used.  A significant difference between the CL and FL is that CL emitted by the shower particles is strongly peaked forward along the shower direction, and falls off rapidly as the shower viewing angle changes.  In contrast, FL is emitted isotropically.  As a result, Cerenkov events are seen only if the shower geometry with respect to the detector is such that the shower is moving towards the detector (viewing angle $\sim10^{\circ}$ or smaller).  Fluorescence events are recorded at all viewing angles.  This fact has a number of consequences for the observation:
\begin{itemize}
\item Cerenkov events are much ``faster'' (total event duration is much shorter) than fluorescence events.
\item At any energy/distance Cerenkov events are more intense (bright) than the fluorescence counterpart, which is why the detector energy threshold is lower for Cerenkov events.
\item The detection volume for Cerenkov events is limited and does not grow with energy the same way it does for fluorescence events.  The shower core must fall in the vicinity of the detector for the viewing angle condition to be met.
\end {itemize}

While the amount of CL emitted at the shower is proportional to the shower size, the amount of light observed at the detector depends strongly on the emission (viewing) angle at which the light is received from the shower.  This results in two significant consequences for the event reconstruction as will be explained in section \ref{sec:recon}:
\begin{itemize}
\item{It is precisely this property which makes the monocular event reconstruction possible.}
\item{It makes the reconstructed shower energy/$x_{max}$ highly sensitive to the error in the reconstructed geometry, and also to the light emission model assumed in the reconstruction}
\end {itemize}

\section{Monocular Event Reconstruction}
\label{sec:recon}

Event reconstruction refers to the determination of the shower geometry with respect to the detector and obtaining a best fit for the shower energy and the depth of shower maximum, $x_{max}$.  It is preferable and simpler to divide this problem into a separate calculation of the best fit shower geometry followed by the calculation of the shower development for the best fit shower energy and $x_{max}$.  Events recorded in monocular mode {\it and} which have short angular track-lengths do not afford the simple division of the reconstruction task, instead they require a combined geometry/profile reconstruction procedure.  Such a procedure, referred to as the Profile Constrained Geometry Fit (PCGF) \cite{AbuZayyad_thesis} was developed for the analysis of HiRes-I data, and was successfully used to produce a significant physics result \cite{Abbasi:2007sv}.

When applied to the TALE Cerenkov events, it was found that the profile constraint method works extremely well.  In particular, it was found that the accuracy of the geometrical reconstruction, section \ref{sec:results}, is comperable with what's expected from a hybrid or a stereo observation.  Noting the accurate reconstruction of the shower zenith angle, a possibility to extend the PCGF method opens up.  By definition, the PCGF precludes obtaining a fit for the shower $x_{max}$.  However, after examining the results from multiple fits with different ``trial'' $x_{max}$ parameters we noticed that the reconstructed geometry was essentially independent from the assumed trial values.  This means that an additional step can be added to the reconstruction procedure in which the PCGF determined geometry can be fixed and a profile/energy fit following standard techniques can be performed.

\section{Results and Discussion}
\label{sec:results}
\begin{figure}[htb]
\centering
\includegraphics[height=1.75in]{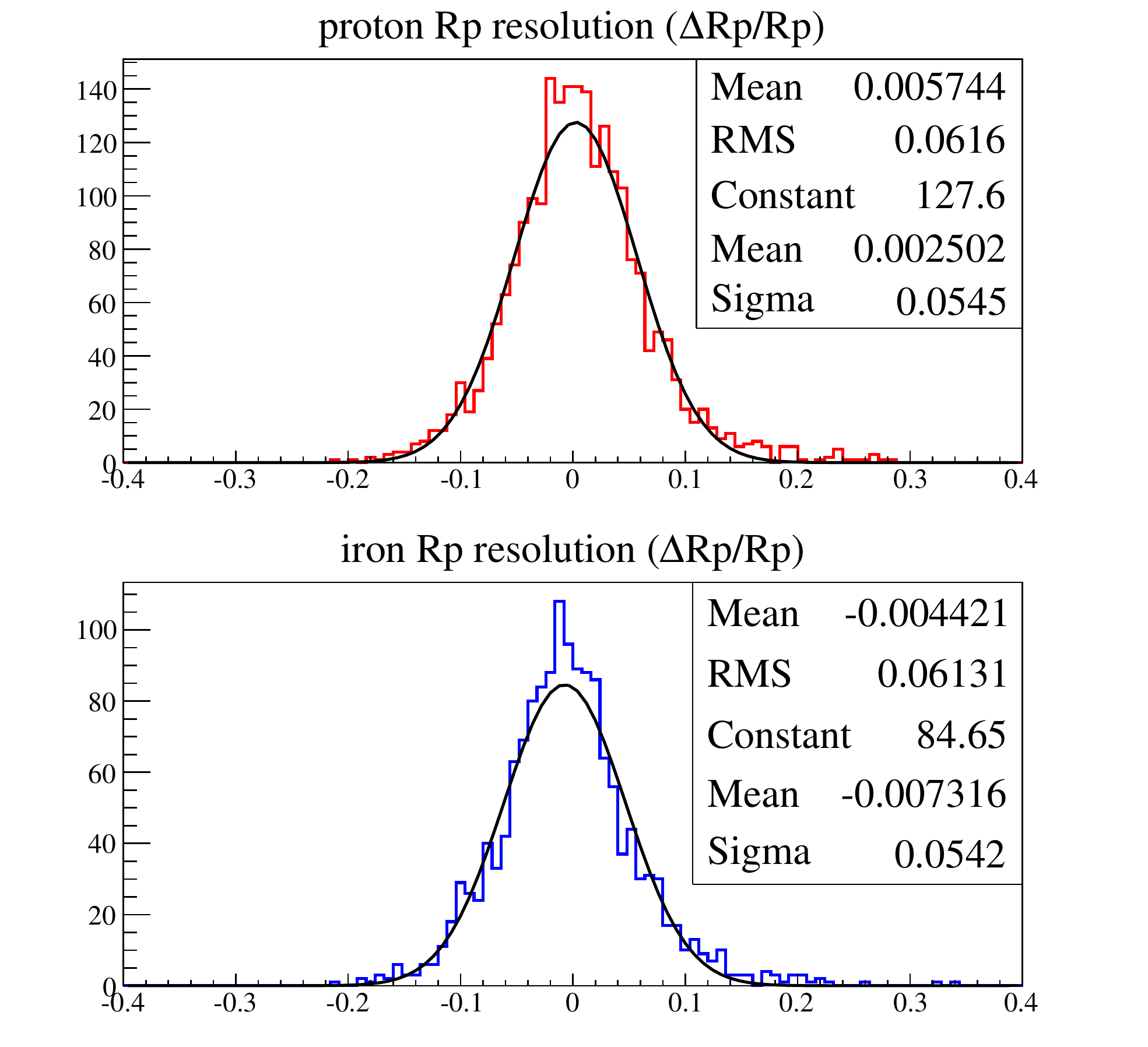}
\includegraphics[height=1.75in]{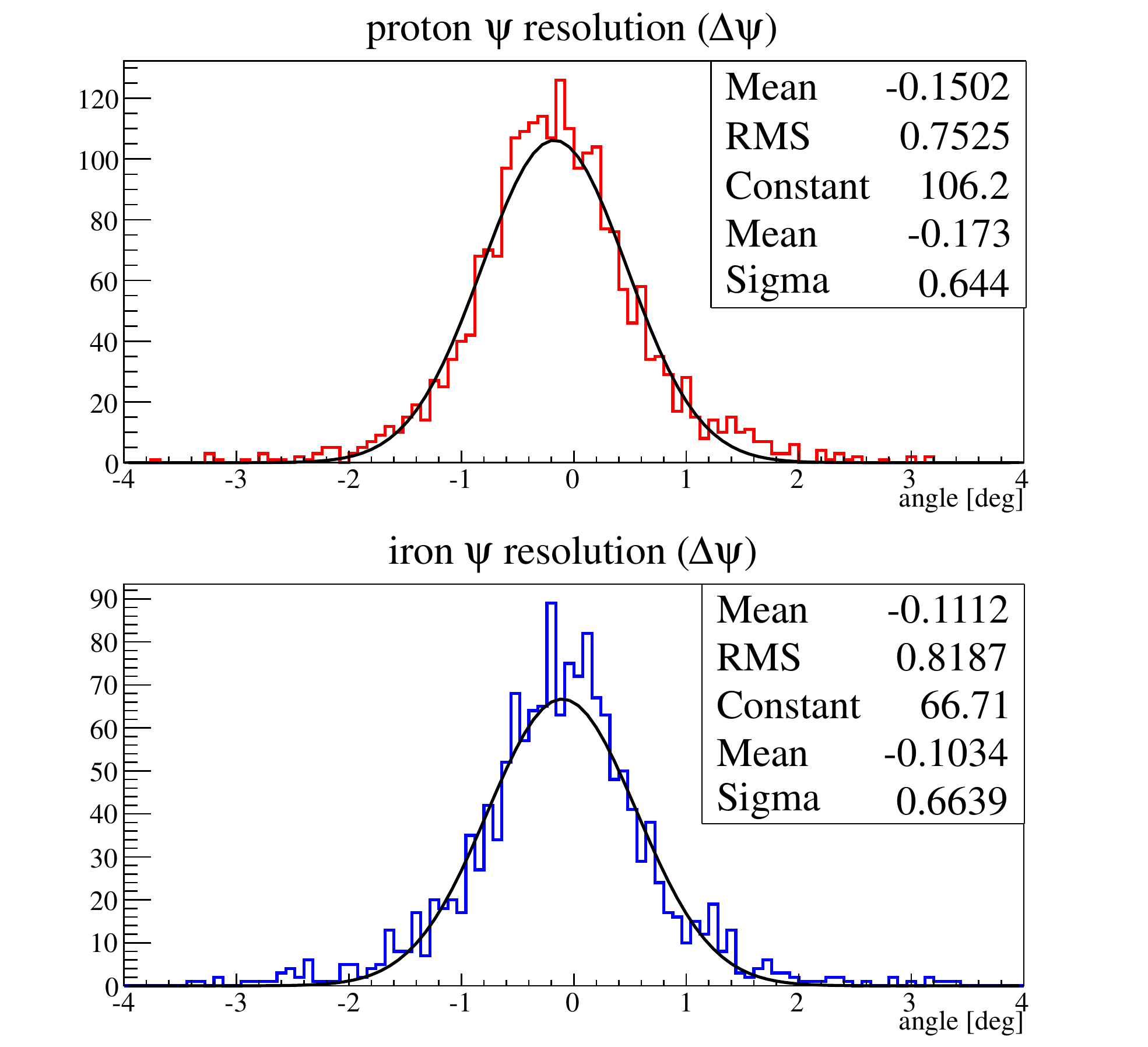}
\includegraphics[height=1.75in]{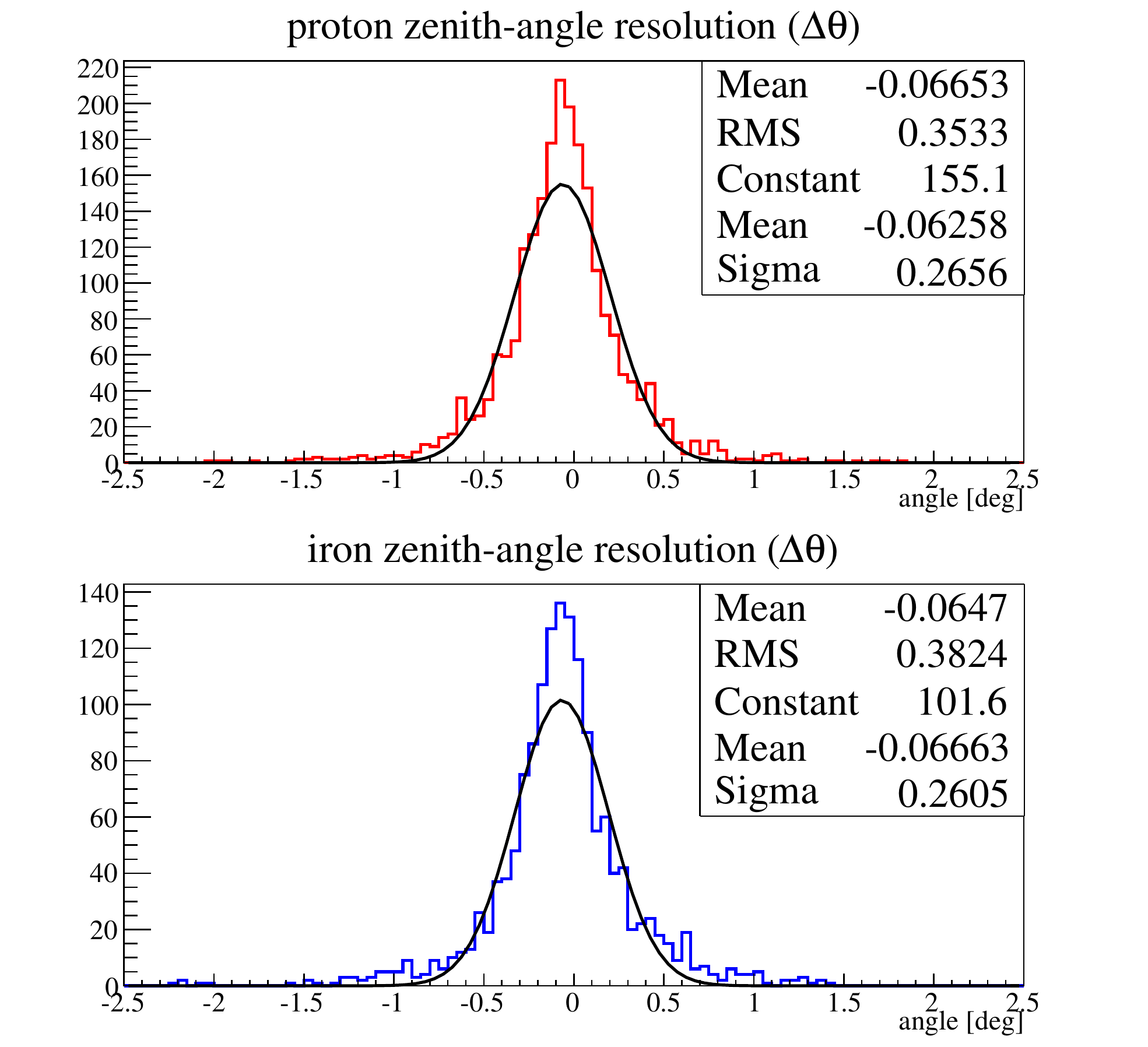}
\caption{Shower Track Geometry Resolution: impact parameter, $R_{p}$ [meters], angle in the plane, $\psi$, and zenith angle, $\theta$, (angles in degrees).  Reconstruction results shown for a MC set generated following a $E^{-2.9}$ energy spectrum.  Only events with reconstructed energies in the range $-2.2 < log_{10}(E [EeV]) < -0.8$ are included.}
\label{fig:resolution}
\end{figure}

As part of the detector response simulation, shower simulation is performed using parametrization of the shower development and light production.  Shower modeling is accurate enough for fluorescence measurements and has been reviewed and updated to include a better description of Cerenkov light production.  Geometrical reconstruction results are shown in Figure \ref{fig:resolution} for data sets of proton and iron showers generated following a $E^{-2.9}$ energy spectrum.  Basic quality cuts (not finalized at the time of this writing) were applied to these sets.  

Further tests using the Corsika/IACT package \cite{Bernlohr:2008kv} confirm these results and their independence of the assumed shower models used in the detector MC and reconstruction code.  The energy and $x_{max}$ reconstruction results will not be shown yet.  We expect to be able to show these results in the coming months.

\section{Acknowledgements}
The Telescope Array experiment is supported 
by the Japan Society for the Promotion of Science through
Grants-in-Aid for Scientific Research on Specially Promoted Research (21000002) 
``Extreme Phenomena in the Universe Explored by Highest Energy Cosmic Rays'', 
and the Inter-University Research Program of the Institute for Cosmic Ray 
Research;
by the U.S. National Science Foundation awards PHY-0307098, 
PHY-0601915, PHY-0703893, PHY-0758342, PHY-0848320, PHY-1069280, 
and PHY-1069286 (Utah) and 
PHY-0649681 (Rutgers); 
by the National Research Foundation of Korea 
(2006-0050031, 2007-0056005, 2007-0093860, 2010-0011378, 2010-0028071, R32-10130);
by the Russian Academy of Sciences, RFBR
grants 10-02-01406a and 11-02-01528a (INR),
IISN project No. 4.4509.10 and 
Belgian Science Policy under IUAP VI/11 (ULB).
The foundations of Dr. Ezekiel R. and Edna Wattis Dumke,
Willard L. Eccles and the George S. and Dolores Dore Eccles
all helped with generous donations. 
The State of Utah supported the project through its Economic Development
Board, and the University of Utah through the 
Office of the Vice President for Research. 
The experimental site became available through the cooperation of the 
Utah School and Institutional Trust Lands Administration (SITLA), 
U.S.~Bureau of Land Management and the U.S.~Air Force. 
We also wish to thank the people and the officials of Millard County,
Utah, for their steadfast and warm support. 
We gratefully acknowledge the contributions from the 
technical staffs of our home institutions as well as 
the University of Utah Center for High Performance Computing (CHPC).

\end{document}

%% file: econfmacros.tex
\def\beq{\begin{equation}}
\def\eeq#1{\label{#1}\end{equation}}
\def\eeqn{\end{equation}}

\def\beqa{\begin{eqnarray}}
\def\eeqa#1{\label{#1}\end{eqnarray}}
\def\eeqan{\end{eqnarray}}

\let\bar=\overbar

\def\Dslash{\not{\hbox{\kern-4pt $D$}}}
\def\dslash{\not{\hbox{\kern-2pt $\del$}}}

\def\msb{{\bar{\ssstyle M \kern -1pt S}}}

